\documentclass[10pt,onecolumn,twoside]{IEEEtran}

\usepackage{amsmath,graphicx,cite}
\usepackage{eurosym}
\usepackage[mathcal]{euscript}
\usepackage{amssymb,amsmath}
\usepackage{algpseudocode}
\usepackage[]{algorithm2e}
\usepackage{verbatim}
\usepackage{graphicx}
\usepackage{longtable}
\usepackage{rotating}
\usepackage{multirow}
\usepackage{tabularx}
\usepackage{bigints}
\usepackage{tikz}
\usepackage{multirow} 
\usetikzlibrary{arrows,backgrounds,snakes}
\usepackage{amsmath,amstext,amsfonts,amssymb}
\usepackage{amsbsy}
\usepackage{amsthm}
\usepackage{diagbox}
\usepackage{mathabx}
\usepackage{color}
\usepackage{float}
\usetikzlibrary{arrows,shapes,positioning,shadows,trees}
\usetikzlibrary{decorations.pathreplacing}

\usepackage{float}
\usepackage{caption}
\usepackage{subcaption}

\usepackage{bbm}

\everymath{\displaystyle}
\title{Analysis of the tradeoff between health and economic impacts of the Covid-19 epidemic}

\author
  {Samson Lasaulce$^\dagger$, Chao Zhang$^{\ddagger,*}$, Vineeth Varma$^\dagger$, and Irinel Constantin Morarescu$^\dagger$
	\thanks{$^\dagger$ CRAN (Universit\'e de Lorraine and CNRS), 54000 Nancy, France.}
\thanks{$^\ddagger$ Central South University, Changsha, China}
\thanks{$^*$Corresponding author: Chao Zhang, zhangchaohust@gmail.com}}
\begin{document}

\maketitle

\begin{abstract}

Various measures have been taken in different countries to mitigate the Covid-19 epidemic. But, throughout the world, many citizens don't understand well how these measures are taken and even question the decisions taken by their government. Should the measures be more (or less) restrictive? Are they taken for a too long (or too short) period of time? To provide some quantitative elements of response to these questions, we consider the well-known SEIR model for the Covid-19 epidemic propagation and propose a pragmatic model of the government decision-making operation. Although simple and obviously improvable, the proposed model allows us to study the tradeoff between health and economic aspects in a pragmatic and insightful way. Assuming a given number of phases for the epidemic (namely, $4$ in this paper) and a desired tradeoff between health and economic aspects, it is then possible to determine the optimal duration of each phase and the optimal severity level (i.e., the target transmission rate) for each of them. The numerical analysis is performed for the case of France but the adopted approach can be applied to any country. One of the takeaway messages of this analysis is that being able to implement the optimal $4-$phase epidemic management  strategy in France would have led to $1.05$ million of infected people and a GDP loss of $231$ billions \euro ~ instead of $6.88 $ millions of infected and a loss of $241$ billions~\euro. This indicates that, seen from the proposed model perspective, the effectively implemented epidemic management strategy is good economically, whereas substantial improvements might have been obtained in terms of health impact. Our analysis indicates that the lockdown/severe phase should have been more severe but shorter, and the adjustment phase occurred earlier. Due to the natural tendency of people to deviate from the official rules, updating measures every month over the whole epidemic episode seems to be more appropriate.
\end{abstract}

\section{Introduction}
\label{sec:introduction}


One of the goals of this work is to provide a simple but exploitable model to measure the quality of the epidemic management strategy implemented by a government to mitigate the health and macro-economic impacts of the Covid-19 epidemic. The quality is measured in terms of the tradeoff between the total number of infected people over a given period of time and the Gross Domestic Product (GDP) loss, under a constraint of the total number of infected people requiring Intensive Care Units (ICU). To reach this objective, we propose a behavioral model for governmental decision-making operations. Although we assume a simple measure for the quality of the lockdown measures and a simple dynamical model (namely, a classical susceptible-exposed-infected-removed (SEIR) model), the proposed approach is seen to be sufficient to constitute a first step into capturing and quantifying the tradeoff under consideration. In contrast with most studies conducted on the Covid-19 epidemic analysis where the primary goal is to refine the SEIR model (see e.g., \cite{peng2020epidemic,lenka-arxiv-2020,victor-ssrn-2020, giordano-arxiv-2020, Roques-Frontiers}) or employ the SEIR model by accounting for local variations (by using a given SEIR model per geographical region - see e.g., \cite{domenico-report-2020}) or for the impact of class type (by age, sex, risk level - see e.g., \cite{khawaja-medrxiv-2020}) our approach is to use the standard SEIR model for an entire country and choose a simple economic model to focus on the study of the tradeoff between health and economic aspects.

Although there have been many several interesting studies on the economic impact of Covid-19 (see e.g., \cite{atkeson2020will,eichenbaum2020macroeconomics, baldwin-book-2020,fernandes-report-2020,mckibbin-report-2020}),the pursued goal of these studies  is not to model the behavior of the government. As a consequence, the proposed tradeoff has not been analyzed, at least formally. In fact, the closest contribution to this direction would be given by \cite{zakary-ijdc} where generic discrete-time epidemics over multiple regions are considered, the particular 4-phase structure is not considered, the focus is not on Covid-19, and the key aspect of the tradeoff analysis is neither developed nor analyzed. Additionally, the numerous studies available on the problem of the transmission rate control generally concern the continuous-time control approach. In this work, the focus is on a multiple phase approach (namely, $4$ phases). In the literature dedicated to epidemic control, one can for instance find that some recent studies on how the lockdown strategies and quarantine can be planned in an optimal fashion \cite{alvarez2020simple, boujakjian2016modeling,casella-arxiv-2020, Rawson-Frontiers}. A common feature to all these works on optimal control and lockdown planning is that the policies under consideration, vary over time in a continuous manner, i.e., the lockdown policy is continuously evolving based on the infected population or just on time. However, from the perspective of a government, implementing such policies is not practical since daily changes of the epidemic control measures are difficult to be implemented and to be followed by people.

\begin{figure}
\begin{center}
 \includegraphics[angle=0,width=12cm]{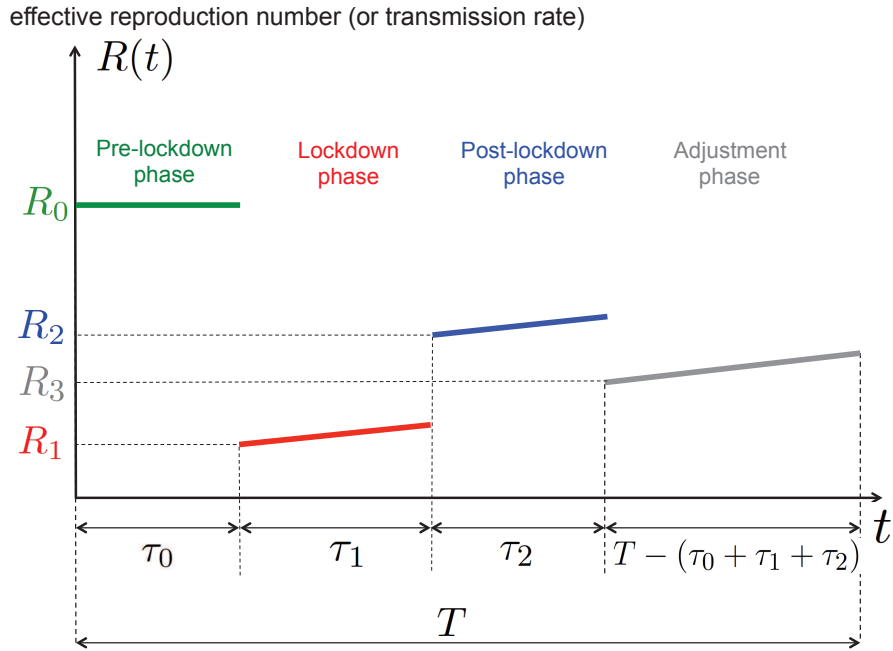}
\caption{One of the goals of this work is to determine numerically, for a given tradeoff between health and economic costs, the best $4-$phase epidemic management policy that is, the best values for $\tau_0$, $\tau_1$, $\tau_2$, $R_1$, $R_2$, $R_3$ (the epidemic time horizon $T$ and the natural reproduction number $R_0$ being fixed).}
\label{fig:R(t)}
\end{center}
\end{figure}

Summarizing, compared to the existing literature on epidemics modelling and control of epidemics, the main contribution of our work is fourfold:
\begin{itemize}
\item a model for capturing the tradeoff between health and economic aspect and therefore for the government decision-making operation is proposed and studied;
\item the focus is on multiple phase control policies and not on general continuous-time control policies (to be precise, $4$ phases are assumed, see Figure~\ref{fig:R(t)});
\item the problem of finding the optimal features of the optimal epidemic management policy (i.e., the target severity level for each phase and the switching time instants) is stated and solved exhaustively. Additionally, to refine the analysis, we assume a simple model for the natural time drift in terms of behavior of people;
\item the numerical analysis of the tradeoff is dedicated to the Covid-19 epidemic and a case study for France.
\end{itemize}


\section{Methods}
\label{sec:model}

\subsection{Epidemic model}

To model the dynamics of the Covid-19 epidemic globally i.e., over an entire country, we assume a standard SEIR model. Let us respectively denote by $s$, $e$, $i$, and $r$ the fractions of the population: being susceptible to be infected by the SARS-Cov2 virus, having been exposed to it, being infected, and being removed (including recoveries and deceases). The epidemic is assumed to obey the following continuous-time dynamics:
\begin{center}
\begin{equation}\label{eq:SIER-system}
\left\{
\begin{array}{ccl}
\displaystyle{\frac{\mathrm{d}s}{\mathrm{d}t}(t)} & = & - \beta(t) i(t)  s(t) \\
&&\\
\displaystyle{\frac{\mathrm{d}e}{\mathrm{d}t}(t)} & = &   \beta(t) i(t)  s(t) - \gamma e(t)\\
&&\\
\displaystyle{\frac{\mathrm{d}i}{\mathrm{d}t}(t)} & = & \gamma e(t) - \delta i(t)\\
&&\\
\displaystyle{\frac{\mathrm{d}r}{\mathrm{d}t}(t)} & = & \delta i(t)\\ 
&&\\
s(t) + e(t) + i(t) + r(t) &=& 1\\
\end{array}
\right.
\end{equation}
\end{center}
where:
\begin{itemize}
\item $\beta(t)$, $t \in \mathbb{R}$, represents the \textbf{time-varying} virus transmission rate;
\item $\gamma$ denotes the rate at which the exposed subject develops the disease (this includes people presenting symptoms and asymptomatics). The period $\frac{1}{\gamma}$ is called the incubation period;
\item $\delta$ denotes the removal rate and $\frac{1}{\delta}$ is called the average recovery period.
\end{itemize}

We assume that the the control action $u(t)$ taken by the decision-maker (the government or possibly a more local decision-maker) has a \textbf{linear effect} on the transmission virus rate. Additionally, the effectiveness of this action is assumed to undergo a non-controllable drift or attenuation effect due to the observed fact that people tend to relax their effort over time \cite{Dagnall-Frontiers}\cite{Anzum-medarxiv}, hence the presence of the attenuation factor $a(t)$ yields:
\begin{equation}\label{eq:beta}
\beta(t) = R_0 \delta - u(t)a(t)
\end{equation}
where $R_0$ is the natural reproduction number (namely, without any control or population awareness), $u(t) \in [0, U]$ is the control action or severity level of the lockdown measures. Note that $U$ corresponds to the most drastic or severe control action (in theory it could reach the value $R_0 \delta$ and make the reproduction number vanishing). In this work, $u(t)$ is a piecewise-constant function. For the numerical analysis, we will assume $a(t) $ to be a linearly decreasing function of time (as detailed in the next section). Therefore, one can define the time-varying effective reproduction number:
\begin{equation}
R(t) = \frac{\beta(t)}{\delta} = R_0 - \frac{u(t)a(t)}{\delta}.
\end{equation}  
As illustrated by Figure~\ref{fig:R(t)}, we are solving an epidemic control problem in which \textcolor{black}{determining the function $u(t)$ or $R(t)$ amounts to jointly determining the switching instants $\tau_0$, $\tau_1$,$\tau_2$} and the targeted reproduction numbers $R_1$, $R_2$, $R_3$; $T$ is a given period of time for the epidemic analysis. In particular, we will determine the best duration of the lockdown phase $\tau_1$ and the corresponding targeted reproduction number $R_1$. Figure~\ref{fig:R(t)} shows for instance that if the lockdown measures taken are such the reproduction number is $R_1$ at the beginning of the lockdown phase, then, because of the drift induced by the typical human behavior, the effective reproduction number increases over time.

\subsection{Time drift or people behavior model}

We propose here a model for the attenuation function $a(t)$, which quantifies the degree to which people relax their effort to implement the government management measures. As the attenuation effect is negligible when a new policy is released, we consider that $a(t)=1$ when $t \in \{\tau_0,\tau_1,\tau_2\}$. The attenuation factor is assumed to increase over time in each phase, and we assume the following piecewise linear behavior between phases:
\begin{equation}\label{eq:attenuation}
a(t)=\left\{
\begin{array}{ccl}
1&&\text{for }  t<\tau_0,\\
&&\\
1- a_{1} ( t-\tau_0 ) &&\text{for } \tau_0\leq t<\tau_1,\\
&&\\
1- a_{2} ( t-\tau_1 ) &&\text{for }   \tau_1\leq t<\tau_2,\\
&&\\
1- a_{3} ( t-\tau_2 ) &&\text{for }  t\geq\tau_2.\\
&&\\
\end{array}
\right.
\end{equation}
where $a_1$, $a_2$, and $a_3$ respectively represent the attenuation coefficients during the lockdown phase, after the lockdown phase, and during the adjustment phase.

\subsection{Decision-maker behavior model}

The proposed model for the behavior of the decision-maker is based on the fact that it wants to obtain a given tradeoff between economic and health aspects. For the cost related to the economics loss, we assume the simplest reasonable model. That is, we assume that economic cost is quadratic in the control action. For the health cost, we assume that it is given by the number of infected people over the given period of time. Therefore, the proposed overall cost consists of a convex combination of these two costs. By minimizing the overall cost, one realizes the desired tradeoff between economic and health aspects. On top of this we impose the number of patients requiring intensive care to be under a given threshold $N_{\max}^{\mathrm{ICU}}$. Thus, the corresponding minimization is performed under a constraint on the number of people infected at any time $t\in[0,T]$: $\sigma N i(t) \leq N_{\max}^{\mathrm{ICU}}$, $N$ being the population size, $N_{\max}^{\mathrm{ICU}}$  the maximum number of ICU patients, and $0 \leq \sigma \leq 1$ is the percentage of infected people requiring intensive care. In France, official records state that the maximum cumulated number of ICU patients has reached $7 \ 148$ (on April 8, 2020) but the capacity over the whole territory has been evaluated to be greater than $15 \ 000$. By denoting $\alpha\in[0,1]$ the weight assigned to the macroeconomic impact of the epidemic and $K_{\mathrm{e}}>0$, $K_{\mathrm{h}}>0$, $\mu >0$ some constants (parameters) defined below, obtaining the desired tradeoff amounts to finding a solution of the following optimization problem (OP) while fixing $\alpha$ to a given value:
{ \begin{equation}\label{eq:OP}
\begin{array}{cl}
\underset{u(t)}{\text{minimize}} & \left\{\begin{split}&\alpha K_{\mathrm{e}} \left\{  \displaystyle{\int_{0}^{\tau_0 + \tau_1} u^2(t) \mathrm{d}t}
 + \frac{1}{\mu_1^2} \displaystyle{\int_{\tau_0 + \tau_1}^{\tau_0 + \tau_1+\tau_2} u^2(t) \mathrm{d}t}+ \frac{1}{\mu_2^2} \displaystyle{\int_{\tau_0 + \tau_1+\tau_2}^{T} u^2(t) \mathrm{d}t}\right\}\\
&+ (1-\alpha) K_{\mathrm{h}} \left[ s(0) - s(T) \right]\end{split}\right\}\\
&\\
\text{subject to}  &  \forall t \in [0,T], \ \sigma N i(t) \leq N_{\max}^{\mathrm{ICU}}  \\
& \tau_1\geq T_{\min}\\
& \text{Equations} \ (\ref{eq:SIER-system}) \ \text{and} \ (\ref{eq:beta})\\
\end{array} 
\end{equation} }
where:
\begin{itemize}
\item $K_{\mathrm{e}}>0$ and $K_{\mathrm{h}}>0$ are constants that weight the economic and health cost functions (they also act as conversion factors allowing one to obtain appropriate units and orders of magnitude);
\item $\tau_0$ and $\tau_1$ represent the lockdown starting time and duration, respectively. \textcolor{black}{$T_{\min}$ is the minimum lockdown duration to make the lockdown policies effective.} The quantity $\tau_2$ represents the duration of the post-lockdown phase;
\item the parameters $\mu_1,\mu_2 \geq 1$ accounts for possible differences in terms of economic impact between the lockdown and post-lockdown phases;
\item $s(0)$ and $s(T)$ are respectively the fractions of the population susceptible at the beginning and the end of the analysis. 
\end{itemize}
We would like to make additional comments concerning the parameter $\mu_1,\mu_2 $. The motivation for introducing $\mu_1,\mu_2$ is twofold. First, after lockdown, people are more aware and act more responsibly than before lockdown. This means that automatic and costless population distancing typically occurs \cite{atkeson2020will,greenstone-workingpaper-2020,andersen-report-2020}. Taking $\mu_1,\mu_2 \geq 1$ precisely amounts to having a smaller reproduction number without any cost for the government. 
{Additionally, as people typically tend to relax their effort to implement the epidemic management measures as time passes, it makes sense to assume that $\mu_1\geq\mu_2$ in our model.}  It also allows one to account for the fact that, after lockdown, the economic activity grows after the lockdown and the effects of the pandemic starts vanishing. This means that, in some sense  we ignore memory effects due to lockdown measures. Further refinements of the proposed model might be considered to account for the lockdown memory effects. This is out of the scope of the present paper but we believe that, this would correspond to assuming $\mu_i <1$.

\subsection{4-Phase optimal control with piecewise constant control actions}

Solving analytically the optimization problem given by (\ref{eq:OP}) is not trivial. However, since we restrict our attention to a certain class of control policies, the problem turns out to be solvable through exhaustive search. 
{Assuming the attenuation factor $a(\tau_0)=a(\tau_1)=a(\tau_2)=1$ (no attenuation at the beginning of each phase) and a constant control action in each phase,} by using the relation $u(t)= \delta [R_0 - R(t)]$, the OP (\ref{eq:OP}) can be rewritten under a more convenient form for numerical purposes:
{ \begin{equation}\label{eq:OP2}
\begin{array}{cl}
\underset{(\tau_0, \tau_1, \tau_2, R_1, R_2, R_3)}{\text{minimize}} & \left\{\begin{split}&\alpha K_{\mathrm{e}} \delta^2 (R_0 - R_1)^2 \tau_1
+ \displaystyle{\frac{\alpha K_{\mathrm{e}} \delta^2  (R_0 - R_2)^2
\tau_2}{\mu_1^2}}+\\
&\displaystyle{\frac{\alpha K_{\mathrm{e}} \delta^2  (R_0 - R_3)^2[T - (\tau_0 + \tau_1+\tau_2)]}{\mu_2^2}}+ (1-\alpha) K_{\mathrm{h}} \left[ s(0) - s(T) \right]\end{split}\right\}\\
&\\
\text{subject to}  &  \forall t \in [0,T], \ \sigma N i(t) \leq N_{\max}^{\mathrm{ICU}}  \\
& \tau_1\geq T_{\min}\\
& R_2>R_1+0.2\\
& \text{Equations} \ (\ref{eq:SIER-system}) \ \text{and} \ (\ref{eq:beta}).\\
\end{array} 
\end{equation} }
{where $R_i$ represents the desired or target reproduction number over Phase $i \in\{1,2,3\}$ without considering the attenuation factor (also, it is the reproduction number at the start of $i$-th phase). The second constraint is introduced here as there is a gap between lockdown reproduction number and after lockdown reproduction number.} Finally, the conversion factors $K_{\mathrm{e}}$ and $K_{\mathrm{h}}$ are chosen as follows. The rationale behind the choice of $K_{\mathrm{e}}$ is that when choosing $\alpha = 1$ the GDP loss should correspond to the best estimations made by economists. The GDP loss over the lockdown period for a given country is denoted by $\Delta \mathrm{GDP}$, the conversion factor $K_{\mathrm{e}}$ is chosen as follows:
\begin{equation}
K_{\mathrm{e}} \delta^2 (R_0 - R_1)^2 \tau_1 = \Delta \mathrm{GDP}.
\end{equation}
For France for example, the GDP loss during the lockdown has been evaluated (on April 20) to be around $120$ billions \euro ~according to the OFCE \cite{ocfe-report-2020}. At last, the constant $K_{\mathrm{h}}$ is merely chosen as $K_{\mathrm{h}}=N$, that is, when $\alpha=0$ the cost function corresponds to the number of people infected over the considered period of time.

\section{Results}
\label{sec:numerical-analysis}

\subsection{General simulation setup}

To perform exhaustive search over the sextuple of variables $(\tau_0, \tau_1, \tau_2, R_1, R_2,R_3)$, time and amplitudes are quantized; we thus use hat notations to indicate corresponding values are quantized. Time is discretized with a step of $24$ hours (that is, one sample for each day) and a time horizon of $\widehat{T}=300$ days (which approximately corresponds to 10 months that is the interval [March 1st, December 31th], for the tradeoff figure presented in Sec 3.4, for computational convenience, we take $\widehat{T}=210$ days corresponding to 7 months from March 1st to September 30th) is assumed. The sets for the possible lockdown starting days, the lockdown duration (in days), post-lockdown phase duration, and the reproduction numbers are as follows: $\widehat{\tau}_0 \in \{1,2,...,30\}$, $\widehat{\tau}_1 \in \{T_{\min},T_{\min}+1,...,90\}$, $\widehat{\tau}_2 \in \{1,2,...,120\}$, $\widehat{R}_1 \in \{0.4, 0.2,..., 1.5\} $, $\widehat{R}_2 \in \{0.4, 0.2,..., 1.5\}$, $\widehat{R}_3 \in \{0.4, 0.2,..., 1.5\}$. {Excluding Figure~\ref{eq:tau1_opt}, \textcolor{black}{due to the physical characteristics of the epidemics in France,} we set $T_{\min}=30$ because the lockdown duration is at least 4 weeks or 1 month to make the lockdown effective in real systems \textcolor{black}{\cite{efficiens-web-2020}\cite{conseil-avis-2020}\cite{SPF}}.} The SEIR model parameters are as follows: $\frac{1}{\delta}=\frac{1}{0.1857} = 5.4$ days, $\frac{1}{\gamma}=\frac{1}{0.16} = 6.25$ days; these choices are consistent with many works and in particular the studies performed for France \cite{salje-sciencemag-2020} and Italy \cite{casella-arxiv-2020}. The population size is set to $N = 66.10^{6}$, the maximum number of patients requiring intensive care is set to $N_{\max}^{\mathrm{ICU}}=15.10^{3}$, and $\sigma = 1.5\%$ \textcolor{black}{\cite{efficiens-web-2020}\cite{conseil-avis-2020}\cite{SPF}}. Notice that this number is only reached for very small values of $\alpha$ (for which the total number of people infected over the analysis duration would be around $9$ millions). The exposed population on March 1 2020 is initialized to $N e(0) = 1.33.10^{5}$. This number is obtained from analyzing the data provided in \cite{efficiens-web-2020}. We consider the number of reported deaths at a given time to be a more reliable way of tracking the evolution of the pandemic rather than the reported number of infected people. \textcolor{black}{Indeed, as soon as one examines absolute values, they typically become irrelevant. For example, during lockdown, because of the lack of tests and measurements, the real number of infected people was much higher than the official number. Now, even when tests were performed intensively, because of false positives, the absolute number of infected were again completely unreliable. For a prevalence of $1/1000$ and a test reliability of $5\%$ of false positives we see that the number of infected for 1000 people is declared to be about 50 whereas the actual number of infected is only 1. Motivated by these critical issues, we have considered figures which are much more reliable such as the number of deceases due to Covid-19 in France \cite{salje-sciencemag-2020}\cite{efficiens-web-2020}. From the number of deaths and the global average rate (worldwide) of the mortality rate (in the range $0.3\%-0.5\%$ when averaged over classes of ages and countries), the reconstructed number of infected people turns out to be more accurate.} Therefore, by fixing the mortality rate to a given value (in \cite{salje-sciencemag-2020} for instance, the mortality rate averaged over all the classes of infected people is evaluated to be around $0.53\%$ for France), one can estimate the exposed and infected population size 3 to 4 weeks before the measured number of deaths due to Covid-19. For our computation, the initial conditions of the ordinary differential equations (ODE) equations are chosen as $s(0)=1-e(0)$, $i(0)=r(0)=0$, and the ODE is solved by using the \textsf{Matlab ode45} solver. Concerning the economic cost for France related to Covid-19, the GDP loss over the lockdown period is estimated by the OFCE \cite{ocfe-report-2020} to be $120$ billions \euro~and we have, as reliable figures, that $\tau_1^{\mathrm{France}}= 55$ days with $R_0^{\mathrm{France}}=3.5$ and $R_1^{\mathrm{France}}=0.6$. We therefore take $K_{\mathrm{e}} = 7.379.10^{9}$ \euro/day. Values of the reproduction number for the first two phases come from past and quite accurate evaluations (see e.g.,\cite{conseil-avis-2020}). The value $R_2^{\mathrm{France}}=0.9 $ is less accurate and corresponds to the assumption that the government has been aiming at giving as much as freedom to the population while avoiding a second wave. Based on available statistics on Covid-19 in France \cite{SPF}, the attenuation coefficients of the drift model have been chosen as follows: $a_1 = 0.1 \%$, $a_2 = 0.2 \%$, $a_3 = 0.2 \%$. Unless stated otherwise, the economic impact parameters are chosen as $\mu_1=1.41$ (i.e., $\mu_1^2 \sim 2$) and $\mu_2=1.3$. Also, when $\alpha$ is assumed to be fixed, it is set to $10^{-4}$. 

\begin{figure}[!h]
\begin{center}
 \includegraphics[angle=0,width=10cm]{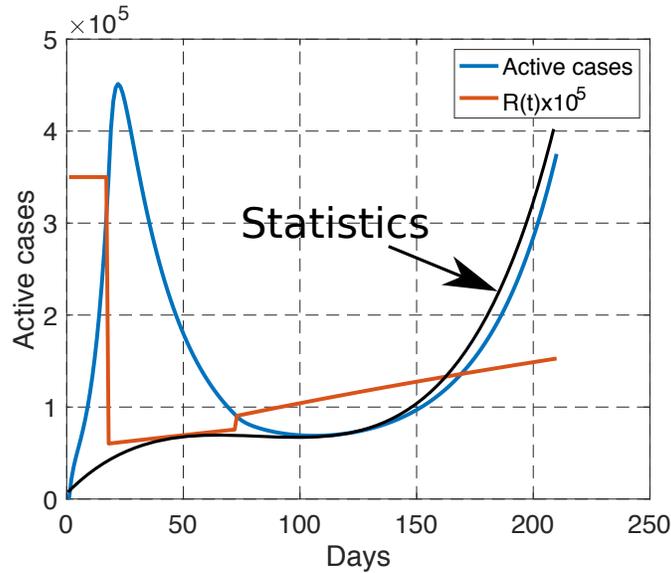}
\end{center}
\caption{Comparison between our model (blue curve) and the reported statistics (black curve) from March 1st to September 30th.\ When the number of tests are sufficiently large, our model matches well with the reported statistics. The reproduction number is linearly increasing  during each phase, and is discontinuous during the transition between phases.}
\label{fig:validation}
\end{figure}

{To justify the choice of the attenuation parameters, that is, $a_1 = 0.1 \%$, $a_2 = 0.2 \%$, $a_3 = 0.2 \%$, we apply the French policy into our model. By comparing the active cases obtained from our model and the official data, it can be illustrated in Fig.~\ref{fig:validation} that our model matches well with the real data, especially in the second-half of the plot, where the number of tests conducted are sufficiently large. This validate the choice of our model and parameters. The mismatch on March and April are mainly due to the lacking number of tests that were taken during the early outbreak of the pandemic, leading to a much lower reported number of active cases.}

\subsection{Optimal tradeoff between economic and health impacts}

\begin{figure}[!h]
\begin{center}
 \includegraphics[angle=0,width=10cm]{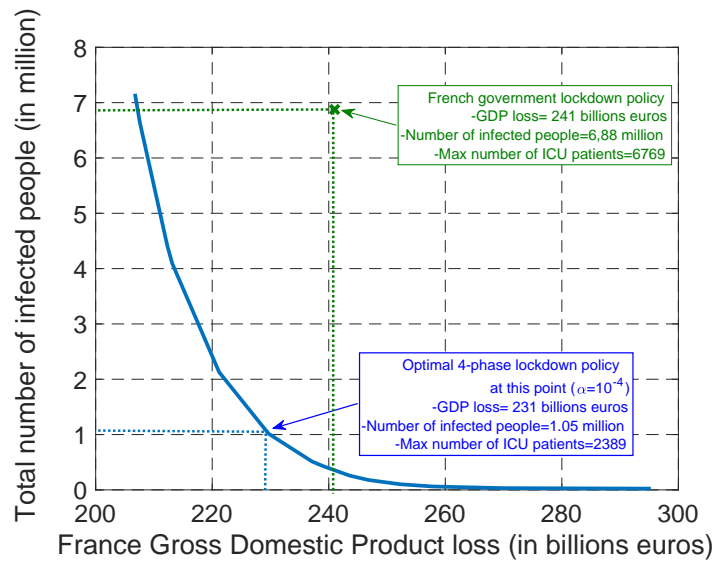}
\end{center}
\caption{The plots represent the possible tradeoffs between health cost (measured in terms of the total number of infected people) and economic cost (measured in terms of GDP loss) that can be obtained (by choosing the best epidemic management policy). In particular, with the assumed model, it is seen, in retrospect, that it would have been possible to divide the number of infected people by about $6$ while saving about $10$ billions \euro~in terms of GDP with the optimal 4-phase strategy.}
\label{fig:tradeoff}
\end{figure}

With the proposed government decision-making model, implementing a desired tradeoff between the health cost and economic cost merely amounts to choosing a given value for the parameter $\alpha$. Figure~\ref{fig:tradeoff} depicts for various values of $\alpha$ in the interval $[10^{-7}, 10^{-4}]$ the total GDP loss and number of infected people that is obtained after choosing the (quantized version of the) sextuple $(\tau_0, \tau_1, \tau_2, R_1, R_2, R_3)$ that minimizes the combined cost given by Equation (\ref{eq:OP2}). At one extreme, when $\alpha$ is relatively large ($\alpha=10^{-4}$) (that is, when the government aims at minimizing the economic cost in the first place - always under the ICU capacity constraint) we see that the best epidemic management strategy leads to a GDP loss over the entire study period [March 1, September 30] is about $206$ billions \euro~with  $7.16$ millions infected, and $15 \ 000$ patients requiring intensive care. At the other extreme, when $\alpha$ is relatively small ($\alpha=10^{-7}$), the GDP loss reaches values as high as $295$ billion \euro~with a total number of newly infected people over the period [March 1, September 30] as low as $23 \ 162$. To evaluate the efficiency of the epidemic management strategy of the French government policy, we have represented the point corresponding to the estimated number of infected people and GDP loss by September 30; with our model, the GDP loss over the period of time of interest is  $241$ billion \euro~and the total number of infected people is about $6.88$ million. What the best tradeoff curve indicates is that there were management policies that would allow the French government to have a better "performance" both in terms of GDP loss and the number of infected people. For instance, we indicate a point for which it would have been possible to have about $1.05$ million people infected (that is, about $6$ times less than what is estimated with the current policy) while ensuring a total GDP loss of $231$ billions \euro. Which type of epidemic management strategy should be used to have such an outcome? The next sections provide a detailed analysis of the features of the optimal strategy.

\subsection{Optimal features of the optimal epidemic management strategy}
\begin{figure}
\begin{center}
 \includegraphics[angle=0,width=10cm]{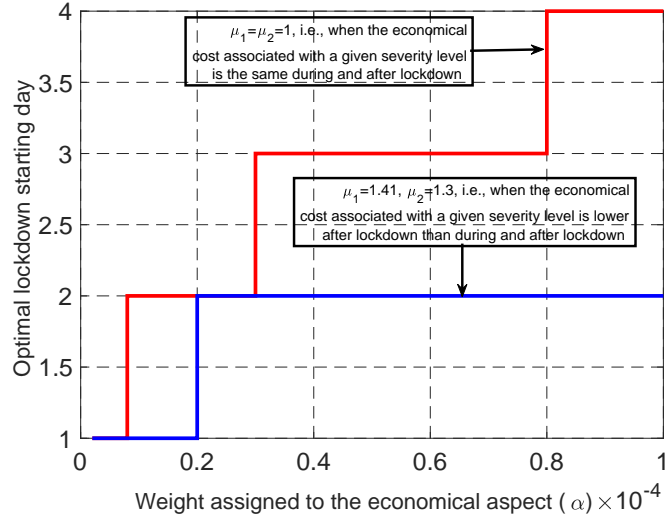}
\end{center}
\caption{When economic losses are assumed to be uniform over time ($\mu=1$), there is some economic incentive to delay the lockdown, but this delay is seen to be at maximum $4$ days. When economic losses are lower after lockdown than during it, it is beneficial to start the lockdown faster, up to $3$ days delay.}
\label{eq:tau0_opt}
\end{figure}

One of the important features for controlling the Covid-19 epidemic which has been well commented in newspapers in various countries is the lockdown starting time. To minimize the health cost, the answer is ready: the lockdown phase should always start as soon as possible. But when one wants to realize a tradeoff between health and economic costs, the answer is less immediate. For different values for $\mu_1$ and $\mu_2$, Figure~\ref{eq:tau0_opt} provides the best day to start locking down the population, for one hundred values of $\alpha$ ranging from $10^{-4} $ to $10^{-6}$. The main message of this figure is that, even for (relatively) large values for $\alpha$ (that is, when the economic cost dominates the health cost), the optimal lockdown starting day should be before March 4th (i.e., $\tau_0\leq4$). This clearly shows that, once an epidemic has been declared, invoking economic damages to delay the lockdown phase is not acceptable. Note that this conclusion holds when economic losses are assumed to be uniform over time ($\mu_1=\mu_2=1$). When the economic cost associated with a given intensity or severity level is lower after lockdown than during it (here $\mu_1=1.41$ and $\mu_2=1.3$), it is always optimal to start locking down as soon as possible. Note that our model does not capture the possible fact that population needs to be psychologically prepared to follow the lockdown measures. In France, by March 17, there were official figures about the epidemic which were sufficiently critical to make the population accept the measures whereas, starting on March 4 (the optimal starting date for $\mu_1=\mu_2=1$) the situation might have not been critical enough to create full adhesion to government measures. 

\begin{figure}
\begin{center}
 \includegraphics[angle=0,width=10cm]{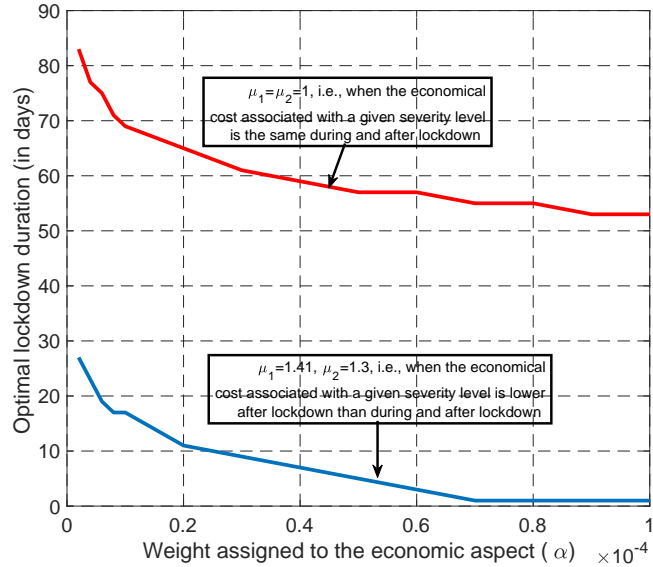}
\end{center}
\caption{When economic losses are assumed to be uniform over time ($\mu=1$), it is optimal to have a long lockdown (typically between $60-80$ days) whatever the tradeoff desired. But, if the economic losses of the third phase are less than during lockdown, the lockdown phase should be shorter.}
\label{eq:tau1_opt}
\end{figure}

A second key feature of the Covid-19 epidemic control strategy was the lockdown phase duration namely, the value of $\tau_1$. \textcolor{black}{To better explore the relationship between $\tau_1$ and $\alpha$ (the tradeoff), we relax the lockdown duration constraint here and set $T_{\min}=1$.} For $(\mu_1,\mu_2)=(1,1)$ and $(\mu_1,\mu_2)=(1.41,1.3)$, Figure~\ref{eq:tau1_opt} provides the optimal lockdown duration (in days) for values of $\alpha$ ranging from $10^{-4} $ to $10^{-6}$. For $(\mu_1,\mu_2)=(1,1)$  (i.e., when economic losses are uniform over time), the optimal duration ranges from $53$ days to $83$ days for a large fraction of the considered interval for $\alpha$. Interestingly, we see that these values are relatively close to the lockdown duration effectively imposed in France namely $55$ days. For larger values of $\mu_1$ and $\mu_2$, the optimal lockdown duration is seen to be much smaller. Therefore, if economic losses are uniform over time, the French government policy seems to be very coherent. On the other hand, if the economic impact is smaller after lockdown, our study suggests shorter lockdown durations. In fact, our results show the existence of a critical value for the tradeoff parameter $\alpha$ above which the second phase of the management of the epidemic should not be present. This means that the optimal control consists of  three phases instead of four.

\begin{figure}
\begin{center}
 \includegraphics[angle=0,width=10cm]{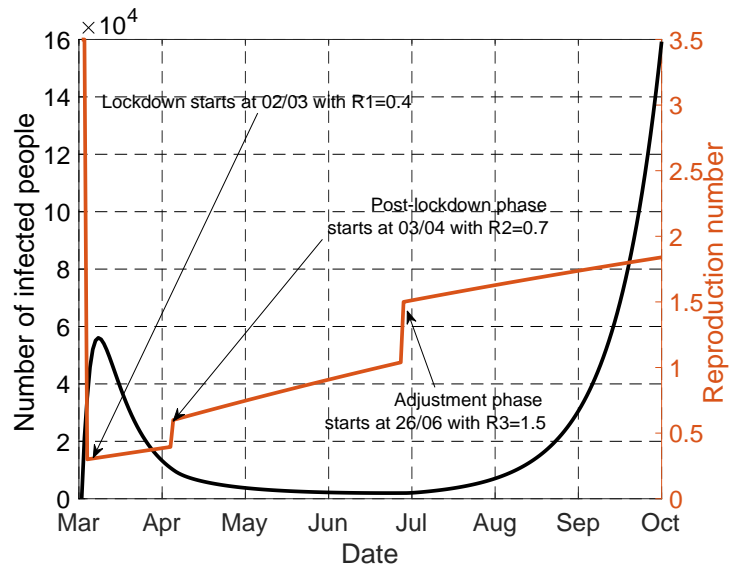}\end{center}
\caption{This figure represents the evolution of number of infected people ($Ni(t)$) and the transmission rate when the optimal policy is adopted.}
\label{fig:4phase_opt}
\end{figure}

To conclude this section, let us consider Figure~\ref{fig:4phase_opt}. For the by default scenario studied in this paper ($(\mu_1,\mu_2) = (1.41, 1.3)$, $\alpha = 10^{-4}$), the figure represents the evolution \textcolor{black}{of number of infected people, that is, $N i(t)$, }and the transmission rate when the optimal policy is adopted. First, it is seen that for the health-economic tradeoff corresponding to $\alpha = 10^{-4}$, there is no interest in delaying the lockdown phase. By "lockdown" phase, the authors mean that it might be any type of phase for which the reproduction number is as low as $0.4$ (versus the estimated $0.6$ in France); very efficient digital tracing and intensive use of face masks is also an option which has been successfully adopted in countries such as South Korea (see e.g., \cite{park-eid-2020}). The optimal lockdown phase duration is seen to be about $1$ month (instead of $2$ for France). We see that the existence of an adjustment phase is part of the optimal policy. For this point, we see that the adjustment phase should have occurred much before in France (end of June versus end of September). The next section is precisely dedicated to the impact of the adjustment phase.  

\textcolor{black}
{
\subsection{Lockdown policies with different $R_0$}
Since the natural reproduction number $R_0$ depends on temperatures, population density, may vary over time due to mutation effects, and in any case is not known perfectly, it is of interest to study the impact of $R_0$ on the obtained characteristics for the optimal epidemic management policy. This is what Fig.~\ref{fig:R0_variable} represents. It is seen that large variations on $R_0$ do not involve large variations on the starting day. For instance, moving from $R_0=2$ to $R_0=3.5$ only changes the optimal date by one day (Day $2$ instead of Day $3$), which confirms the need to act fastly even when the transmission is more limited (e.g., thanks to higher temperatures or lower population density). Note that this holds even if the economic impact is accounted for. It is also good for economical aspects to react fastly to an epidemics. For the optimal reproduction number during lockdown (namely, $R_1$) it is also seen that moving a scenario in which $R_0=2$ to $R_0=3.5$ does not change very significantly the results: the target severity (or freedom) level would correspond to $R_1=0.4$ instead $R_1=0.6$, which shows that the severity should be high during lockdown even in countries or regions where transmission is more limited.     
\begin{figure}[!h]
\begin{center}
 \includegraphics[angle=0,width=10cm]{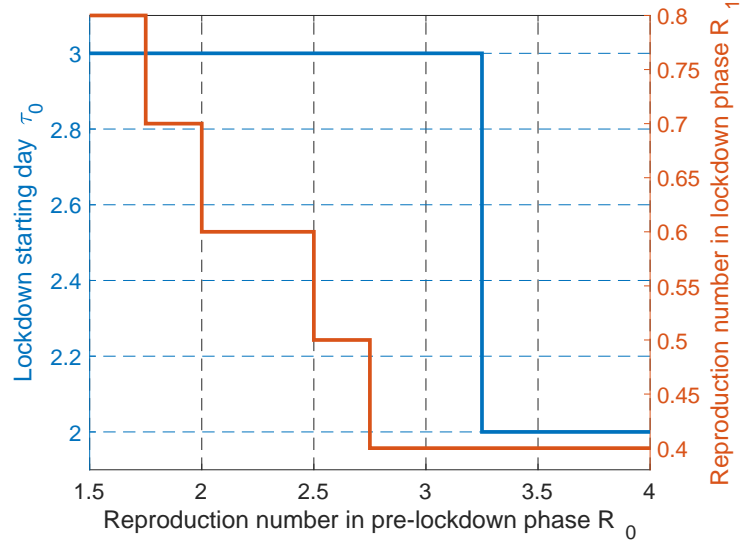}
\end{center}
\caption{\textcolor{black}{Comparison of lockdown policies with different $R_0$. When the situation is worse, a sooner and more strict lockdown strategy should be applied.}}
\label{fig:R0_variable}
\end{figure}
}

\begin{figure}[!h]
\begin{center}
 \includegraphics[angle=0,width=10cm]{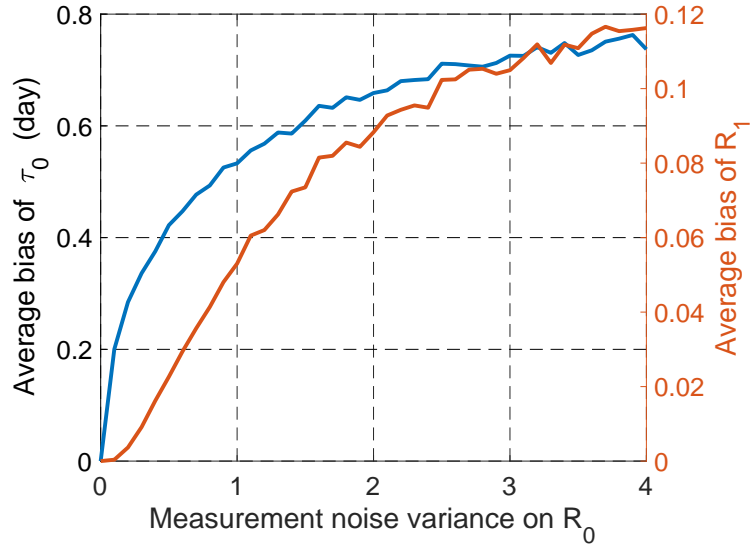}\end{center}
\caption{\textcolor{black}{Influence of uncertainties on $R_0$ on the optimal values for $\tau_0 $ and $R_1$.}}
\label{fig:R0_uncertainty}
\end{figure}

\textcolor{black}{Alternatively, the impact of $R_0$ uncertainty can be evaluated by adding a perturbation on $R_0$. It is assumed that what is known to determine the optimal parameters is $\widehat{R}_0 = R_0 + \Delta$, $\Delta$ being a Gaussian noise $\Delta\sim\mathcal{N}(0,\sigma^2)$. The reproduction number have to stay in a given interval of physical relevance of the form $[R_{\min},R_{\max}]$. Thus the noise is imposed to stay in the interval $[-R_0+R_{\min},-R_0+R_{\max}]$. With $R_0=3.5$, $R_{\min}=1$ and $R_{\max}=4$, Fig.~\ref{fig:R0_uncertainty} depicts the average bias for $\tau_0$ and $R_1$ induced by uncertainty on $R_0$. The average biases for $\tau_0$ and $R_1$ are defined by $\mathbb{E}_{\Delta}[|\widehat{R}_1-R_1|]$ and $\mathbb{E}_{\Delta}[|\widehat{\tau}_0-\tau_0|]$, where $\widehat{R}_1$ and $\widehat{\tau}_0$ are obtained with the noisy reproduction number $\widehat{R}_0$. Remarkably, the impact of the corresponding noise on the results is seen to be very reasonable and does not affect the main conclusions drawn in the first version of the paper. This indicates that the conducted analysis is robust against some forms of uncertainties. But of course, as mentioned previously, a deeper analysis would be required to state more general conclusions.}

{In addition, a more global sensitivity analysis to other SEIR model parameters, i.e, $\delta$ and $\gamma$, have been conducted by simulations.  A similar conclusion can be drawn. This shows that our conclusions are quite robust to effects such as parameter uncertainties. This is in part due to the fact that we are mostly seeking discrete parameters and not continuous parameters, which creates a form of robustness.}

\subsection{Impact of the adjustment phase}

\begin{figure}
\begin{center}
 \includegraphics[angle=0,width=10cm]{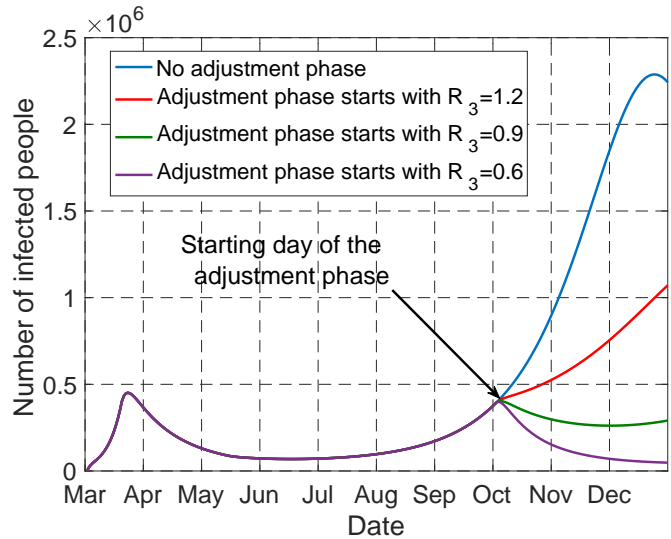}
\end{center}
\caption{The figure shows the impact of the adjustment phase. Only by imposing $R(t) < 1.2$ over this phase, the number of people admitted in ICU does not exceed the capacity of France by the end of 2020.}
\label{fig:reinforcing}
\end{figure}

Always for the typical scenario presented in the general setting part, Figure~\ref{fig:reinforcing} depicts the evolution of the \textcolor{black}{number of infected people} (in France) for the policy effectively implemented over the period [March 1, September 30]. Here, only the adjustment phase is assumed to be optimizable. The figure allows one to quantify the impact of the severity level of the adjustment phase. Without the adjustment phase, the fraction of infected people is such that the number of people requiring intensive care exceeds the double of the maximum ICU capacity of France. \textcolor{black}{However, by implementing measures such that $R(t) < 1.2$ over the adjustment phase, the constraint on the ICU capacity is not violated by the end of 2020. Furthermore, to avoid the overwhelming health service for a longer time, it is better to implement  measures such that $R(t) < 1$ over the adjustment phase since the fraction of infected people will be non-increasing with $R(t) < 1$.}

\section{Discussion}
\label{sec:conclusion}

In this work, we propose to model the behavior of a government as far as the epidemic control is concerned. The proposed model, despite its simplicity, has the merit of being able to capture the fundamental tradeoff between economic and health aspects. Obviously, to capture other effects such as the psychological effects of measures on people, a more general model should be considered. The proposed model allows one to provide quantitative answers to issues which have been well commented in the media. For example, even when a government chooses to assign a high value to the economic aspect, it is seen that the best strategy is almost always to implement a severe phase as soon as possible. This severe phase involves locking the population down, as most countries did, or to make intensive use of digital tracing and face masks as South Korea did. In the latter case, a loss in terms of privacy is the price to be paid for having more freedom. When inspecting the obtained numerical results performed for France, it is seen that the optimal features for the lockdown/severe phase require targeting a \textbf{reproduction number smaller} than the one achieved in France ($0.4$ vs $0.6$), while having a \textbf{shorter duration for the severe phase} ($1$ month instead of $2$). Note that some countries adopted measures which were more severe than France. For instance, China has been imposing the use of a given food supply system which was very efficient in terms of epidemic mitigation \cite{wu-report-2020, lau-jtm-2020, fang-report-2020}. Then, by planning an adjustment phase at the right time, we have seen that as a final result, the number of infected people can be reduced by a factor of $6$ when compared to the current French policy ($1.05$ million of infected instead of $6.88$ millions for the current policy) while having similar GDP losses ($231$ billions \euro ~instead of $241$ billions \euro). We have seen that by considering a simple model as we have studied, the need for an adjustment phase could have been anticipated. Such a phase is necessary to avoid the number of patients under intensive care exceeding the capacity of the ICUs. Also, because of the natural tendency of  humans to deviate from rules over time, it appears that measures should be \textbf{updated about every month} and not less frequently. 
 
\section*{Author Contributions}

SL, CZ, VV, and CM designed the proposed model. SL and CZ conducted the review of the related literature. SL collected the data to calibrate the assumed model. CZ and SL performed the simulations. All authors contributed to the writing of the manuscript.

\section*{Funding}
This work was partially supported by ANR via the grant
NICETWEET, number ANR-20-CE48-0009.

\section*{Acknowledgments}
The authors acknowledge the support of the French National Research Agency ANR and the French National Research Center CNRS. Part of this manuscript has been released as a pre-print at medRxiv \cite{lasaulce-medRxiv-2020}.


\section*{Data Availability Statement}
Publicly available datasets were analyzed in this study. This data can be found in the following website: \textit{www.santepubliquefrance.fr.}

\bibliographystyle{unsrt}

\begin{thebibliography}{10}
 \bibitem{peng2020epidemic} 
 Peng L, Yang W, Zhang D, Zhuge C, Hong L. Epidemic analysis of COVID-19 in China by dynamical modeling. arXiv preprint arXiv:2002.06563. 2020 Feb 16.  
 \bibitem{lenka-arxiv-2020}
Pribylova L, Hajnova V. SEIAR model with asymptomatic cohort and consequences to efficiency of quarantine government measures in COVID-19 epidemic. arXiv preprint arXiv:2004.02601. 2020 Apr 6.
 \bibitem{victor-ssrn-2020}
 Victor A. Estimation of the probability of reinfection with COVID-19 coronavirus by the SEIRUS model. Available at SSRN 3571765. 2020 Apr 8.
 \bibitem{giordano-arxiv-2020}
Giordano G, Blanchini F, Bruno R, Colaneri P, Di Filippo A, Di Matteo A, Colaneri M. A SIDARTHE model of COVID-19 epidemic in Italy. arXiv preprint arXiv: 2003.09861. 2020 Mar 22.
 \bibitem{Roques-Frontiers}
 Roques L, Klein EK, Papa\"{i}x J, Sar A, Soubeyrand S. Impact of Lockdown on the Epidemic Dynamics of COVID-19 in France. Frontiers in Medicine. 2020 Jun 5;7:274. doi: 10.3389/fmed.2020.00274
 \bibitem{domenico-report-2020}
 Di Domenico L, Pullano G, Pullano G, Hens N, Colizza V. Expected impact of school closure and telework to mitigate COVID-19 epidemic in France. COVID-19 outbreak assessment. EPIcx Lab. 2020;15.
 \bibitem{khawaja-medrxiv-2020}
 Khawaja AP, Warwick AN, Hysi PG, Kastner A, Dick A, Khaw PT, Tufail A, Foster PJ, Khaw KT. Associations with Covid-19 hospitalisation amongst 406,793 adults: the UK Biobank prospective cohort study. medRxiv. 2020 Jan 1. doi: 10.1101/2020.05.06.20092957
 \bibitem{atkeson2020will}
 Atkeson A. What will be the economic impact of covid-19 in the US? Rough estimates of disease scenarios. National Bureau of Economic Research; 2020 Mar 19. doi: 10.3386/w26867
 \bibitem{eichenbaum2020macroeconomics}
Eichenbaum MS, Rebelo S, Trabandt M. The macroeconomics of epidemics. National Bureau of Economic Research; 2020 Mar 19. doi: 10.3386/w26882
 \bibitem{baldwin-book-2020}
 Baldwin R, Weder di Mauro B. Economics in the Time of COVID-19.
 \bibitem{fernandes-report-2020}
 Fernandes N. Economic effects of coronavirus outbreak (COVID-19) on the world economy. Available at SSRN 3557504. 2020 Mar 22. doi: 10.2139/ssrn.3557504
 \bibitem{mckibbin-report-2020}
 McKibbin WJ, Fernando R. The global macroeconomic impacts of COVID-19: Seven scenarios. doi: 10.2139/ssrn.3547729
 \bibitem{zakary-ijdc}
 Zakary O, Rachik M, Elmouki I. On the analysis of a multi-regions discrete SIR epidemic model: an optimal control approach. International Journal of Dynamics and Control. 2017 Sep 1;5(3):917-30. doi: 10.1007/s40435-016-0233-2
 \bibitem{alvarez2020simple}
 Alvarez FE, Argente D, Lippi F. A simple planning problem for covid-19 lockdown. National Bureau of Economic Research; 2020 Apr 9. doi: 10.3386/w26981
 \bibitem{boujakjian2016modeling}
 Boujakjian H. Modeling the spread of Ebola with SEIR and optimal control. SIAM Undergraduate Research Online. 2016 Jun 27;9:299-310.
 \bibitem{casella-arxiv-2020}
 Casella F. Can the COVID-19 epidemic be managed on the basis of daily data?. arXiv preprint arXiv:2003.06967. 2020 Mar 16.
 \bibitem{Rawson-Frontiers}
 Rawson T, Brewer T, Veltcheva D, Huntingford C, Bonsall MB. How and when to end the COVID-19 lockdown: an optimization approach. Frontiers in Public Health. 2020 Jun 10;8:262. doi: 10.3389/fpubh.2020.00262
 \bibitem{Dagnall-Frontiers}
 Dagnall N, Drinkwater KG, Denovan A, Walsh RS. Bridging the gap between UK Government strategic narratives and public opinion/behavior: Lessons from COVID-19. Frontiers in Communication. 2020 Sep 17;5. doi: 10.3389/fcomm.2020.00071
 \bibitem{Anzum-medarxiv}
 Anzum R, Islam MZ. Mathematical Modeling of Coronavirus Reproduction Rate with Policy and Behavioral Effects. medRxiv. 2020 Jan 1. doi: 10.1101/2020.06.16.20133330
 \bibitem{greenstone-workingpaper-2020}
 Greenstone M, Nigam V. Does social distancing matter?. University of Chicago, Becker Friedman Institute for Economics Working Paper. 2020 Mar 25(2020-26). doi: 10.2139/ssrn.3561244
 \bibitem{andersen-report-2020}
 Andersen M. Early evidence on social distancing in response to COVID-19 in the United States. Available at SSRN 3569368. 2020 Apr 6. doi: 10.2139/ssrn.3569368
 \bibitem{ocfe-report-2020}
 Observatoire fran\c{c}ais des conjonctures~\'{e}conomiques (\uppercase{OFCE}).
\newblock \'{E}valuation au 20 avril 2020 de l'impact \'{e}conomique de la
  pand\'{e}mie de \uppercase{COVID-19} et des mesures de confinement en
  {F}rance.
\newblock {\em OFCE policy brief}, 2020.
 \bibitem{salje-sciencemag-2020}
 Salje H, Kiem CT, Lefrancq N, Courtejoie N, Bosetti P, Paireau J, Andronico A, Hoz\'{e} N, Richet J, Dubost CL, Le Strat Y. Estimating the burden of SARS-CoV-2 in France. Science. 2020 May 13. doi: 10.1126/science.abc3517
 \bibitem{efficiens-web-2020}
 Open~Stats Coronavirus.
\newblock \uppercase{C}ovid-19 statistiques/\uppercase{F}rance.
\newblock Technical report,
  https://www.coronavirus-statistiques.com/stats-globale/covid-19-par-pays-nombre-de-cas/,
  2020.
 \bibitem{conseil-avis-2020}
 Delfraissy JF, Atlani-Duault L, Benamouzig D, Bouadma L, Casanova JL et al.
\newblock Sortie progressive de confinement: Pr\'{e}requis et mesures phares.
\newblock {\em \uppercase{C}onseil scientifique {C}ovid-19}, 2020.
 \bibitem{SPF}
 Sant\'e \uppercase{P}ublique \uppercase{F}rance.
\newblock {\em www.santepubliquefrance.fr}.
 \bibitem{park-eid-2020}
 Park YJ, Choe YJ, Park O, Park SY, Kim YM, Kim J, Kweon S, Woo Y, Gwack J, Kim SS, Lee J. Contact tracing during coronavirus disease outbreak, South Korea, 2020. Emerging infectious diseases. 2020 Oct;26(10):2465-8. doi: 10.3201/eid2610.201315
 \bibitem{wu-report-2020}
 Wu Z, McGoogan JM. Characteristics of and important lessons from the coronavirus disease 2019 (COVID-19) outbreak in China: summary of a report of 72 314 cases from the Chinese Center for Disease Control and Prevention. Jama. 2020 Apr 7;323(13):1239-42. doi: 10.1001/jama.2020.2648
 \bibitem{lau-jtm-2020}
 Lau H, Khosrawipour V, Kocbach P, Mikolajczyk A, Schubert J, Bania J, Khosrawipour T. The positive impact of lockdown in Wuhan on containing the COVID-19 outbreak in China. Journal of travel medicine. 2020. doi: 10.1093/jtm/taaa037
 \bibitem{fang-report-2020}
 Fang H, Wang L, Yang Y. Human mobility restrictions and the spread of the novel coronavirus (2019-ncov) in china. National Bureau of Economic Research. 2020 Mar 27. doi: 10.3386/w26906
 \bibitem{lasaulce-medRxiv-2020}
 Lasaulce S, Varma VS, Morarescu C, Siying L. How efficient are the lockdown measures taken for mitigating the Covid-19 epidemic?. medRxiv. 2020 Jun 4. doi: 10.1101/2020.06.02.20120089

\end{thebibliography}

\end{document}